\documentclass{nature}

\title{Sub-cycle time resolution of multi-photon momentum transfer in strong-field ionization} 

\author{Benjamin Willenberg$^{1}$, Jochen Maurer$^{1}$, Benedikt W. Mayer$^{1}$ \& Ursula Keller$^{1}$ }

\usepackage{graphicx}
\makeatletter
\let\saved@includegraphics\includegraphics
\AtBeginDocument{\let\includegraphics\saved@includegraphics}
\renewenvironment*{figure}{\@float{figure}}{\end@float}
\makeatother

\usepackage[utf8]{inputenc}
\usepackage[T1]{fontenc}

\usepackage{color}
\begin{document}

\maketitle

\begin{affiliations}
 \item Department of Physics, ETH Zurich, 8093 Zurich, Switzerland
\end{affiliations}

\begin{abstract}
During multi-photon ionization of an atom it is well understood how the involved photons transfer their energy to the ion and the photoelectron.
However, the transfer of the photon linear momentum is still not fully understood.
Here, we present a time-resolved measurement of linear momentum transfer along the laser pulse propagation direction.
Beyond the limit of the electric dipole approximation we observe a time-dependent momentum transfer.
We can show that the time-averaged photon radiation pressure picture is not generally applicable and the linear momentum transfer to the photoelectron depends on the ionization time within the electromagnetic wave cycle using the attoclock technique. 
We can mostly explain the measured linear momentum transfer within a classical model for a free electron in a laser field. 
However, corrections are required due to the interaction of the outgoing photoelectron with the parent ion and due to the initial momentum when the electron appears in the continuum. 
The parent ion interaction induces a measurable negative attosecond time delay between the appearance in the continuum of the electron with minimal linear momentum transfer and the point in time with maximum ionization rate. 
\end{abstract}

\parindent 0pt
\emph{The main idea and first results of this work have been presented in \cite{Willenberg_17, Willenberg_18, Maurer_18 }. }

Photon linear momentum transfer, i.e. momentum transfer along the laser beam propagation axis, upon the interaction of light with matter is one of the most fundamental processes in physics. It impacts a broad range of scientific fields, ranging from laboratory-scale photoionization experiments \cite{Dorner_2000} to plasma physics \cite{Esirkepov_2004, Pegoraro_2007} and laser cooling of microscopic \cite{Wineland_1978, Aspect_1988} and macroscopic objects \cite{Gigan_2006} and it is the underlying mechanism for the occurrence of radiation pressure.

The simplest example  for a process that involves transfer of linear momentum from a photon to an electron is Compton-scattering, where a photon scatters from a free electron. Whereas the fundamental concepts of energy and momentum conservation forbid the complete absorption of the photon by the free electron \cite{Landau_4}, photons can be absorbed by a bound electron during photoionization.
 
In the case of single photon ionization, the linear momentum of the photon ${E_{ph}}/{c}$ with the photon energy $E_{ph}$ ($c$ denotes the speed of light) is transferred to the electron-ion system along the laser propagation direction. For sufficiently high photon energies, basically the complete linear momentum is transferred to the outgoing electron \cite{Sommerfeld_1930, schiff1955quantum}. At low ionization potentials, the electron momentum can even exceed ${E_{ph}}/{c}$ \cite{Michaud_1970, Seaton_1995, Massacrier_1996, Chelkowski_2014}.
In this case the ion receives a momentum in the opposite direction
\cite{Sommerfeld_1930, Michaud_1970, Seaton_1995, Massacrier_1996, Chelkowski_2014}.

The situation changes drastically if we consider photon energies well below the ionization potential of the target and high laser intensities. In this case, multiple photons are involved in the ionization process.
So far, studies on linear momentum sharing and transfer from the laser field to ions and photoelectrons dealt mainly with the time-averaged final photoelectron momenta \cite{Moore_1995, Smeenk_2011, Ludwig_2014, Chelkowski_2014, Chelkowski_2015} and with the ionization-phase-dependent momentum transfers upon recollision \cite{Liu_2013, Maurer_2018}. 
However, to the best of our knowledge, there has been no experimental study on the time-dependent linear momentum transfer during photoionization -- neither for single-photon nor for multi-photon ionization processes.

\begin{figure}
\includegraphics{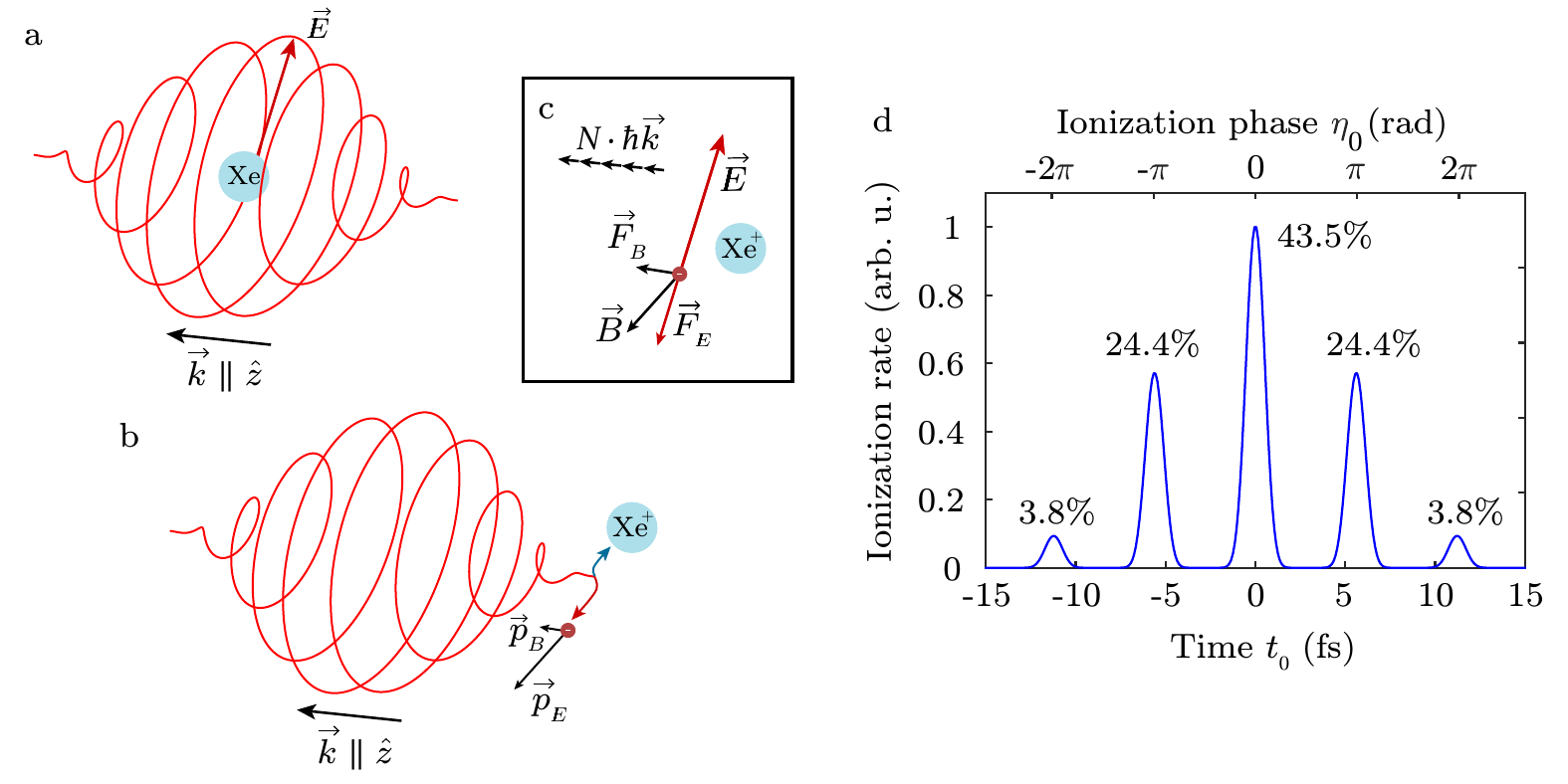}
\caption{\linespread{1.2}\selectfont{}
    \textbf{Timing of linear momentum transfer probed with the attoclock method.}
The rotating electric field vector serves as a time reference to clock the linear momentum transfer onto the electron.
    ({\bf a}) The xenon atom at the time of ionization when the electric field is maximal. 
    ({\bf b}) The electron after the pulse has passed. The electron leaves the pulse at an angle of $\sim 90^\circ\, (=\pi/2)$ with respect to the electric field vector at the time of ionization. Deviations from this angle are caused by the influence of the parent-ion interaction and ionization delay times.
    ({\bf c}) Illustration of the forces acting on the outgoing electron after it was released from the ion. The magnetic field component of the laser field is responsible for a force in laser beam propagation direction. This force can be understood in terms of transfer of linear photon momentum onto the outgoing electron.  
    ({\bf d}) Ionization rate based on the ADK-theory \cite{ADK} during the pulse as function of the ionization phase $\eta_0$ as well as the time $t_0$.}
\label{Fig1}
\end{figure}

Here, we present the first study on the time-resolved linear momentum transfer in strong-field ionization. We achieve sub-cycle time resolution on an attosecond scale by employing the attoclock method \cite{Eckle_2008a, Eckle_2008b}. In this technique, the rotating electric field vector serves as reference for the timing of ionization processes on an attosecond time scale. The measurement method is illustrated in Fig.~\ref{Fig1}. 
In a semi-classical picture the photoelectron is released to the continuum around the peak of the laser electric field (Fig.~\ref{Fig1} (a)). During the subsequent interaction with the electromagnetic pulse the electron is accelerated by the Lorentz force $\vec F_L = -e \cdot (\vec E + \vec v \times \vec B)$ (Fig.~\ref{Fig1} (b)) leading to a final momentum of the electron after the pulse with a component in the polarization plane $\vec p_\perp$ and in laser beam direction $p_z$ (Fig.~\ref{Fig1} (c)). In the multiphoton picture of the ionization process the $p_z$-drift of the electron corresponds to a partial transfer of photon linear momentum.
Although strong-field ionization is in principle a multi-cycle process, the attoclock method allows us to access the dynamics within a cycle. In the polarization plane, the streaking angle reflects the ionization time within a laser optical cycle (see supplementary information).
The contribution from different cycles is shown for the case of zero carrier envelope phase \cite{Telle1999} in Fig.~\ref{Fig1} (d). For our experimental parameters the main contributions stem from the central cycle (43.5 \%) and the neighbouring cycles (24.4 \% each).

To understand the physics of linear momentum transfer in multi-photon strong-field ionization let us first consider a free electron that interacts with light.
If the number of photons involved in the process is sufficiently high, the laser field can be described by a classical electromagnetic field. A widely used model for this situation is based on the classical theory of the high-intensity Thomson scattering, i.e.~the low-energy limit of Compton scattering \cite{Sarachik_1970}: Governed by the laws of classical mechanics, the electron gets accelerated by the electric field of the light. 
If a free electron is passed by an intense classical light pulse, its final momentum is equal to its initial momentum once the laser pulse has completely vanished again. 
However, in the case when the free classical electron is born with an initial momentum $\vec{p}_0$ during the classical pulse, the laser field transfers drift kinetic energy to the electron. 
The dominant fraction of the momentum of the electron after the pulse is $\propto -\vec{A(\eta_0)}$ and directed in the polarization plane, i.e. the plane perpendicular to the propagation direction of the laser pulse. Only a small fraction of the transferred momentum points along the beam propagation direction $z$. 

The sudden appearance of a classical electron in the classical field of the laser pulse is one of the assumptions in the widely used semiclassical two-step models of strong-field ionization: The electron leaves the bound state with essentially zero momentum along the instantaneous electric field direction and is subsequently accelerated by the laser field \cite{TwoStep}. 
The initial position and momentum of the electron are based on the laws of quantum mechanics.
If the parent-ion interaction is neglected, the final momentum of the ionized photoelectron with the initial momentum $\vec{p}_0$ at a phase $\eta_0$ of the laser field can be calculated analytically: $\vec{p}_{f,\perp} = \vec{p}_{0,\perp}-\vec{A}(\eta_0)$ is the final momentum component in the polarization plane, where $\vec{p}_{0,\perp}$ denotes the electron's initial momentum in the polarization plane and $\vec{A}(\eta_0)$ the vector potential at the phase $\eta_0$ (atomic units are used throughout). 
In the propagation direction of the light, $z$, the final momentum is governed according to classical physics in the non-relativistic limit, however without the electric dipole approximation, by
\begin{eqnarray}
\label{eqn:Bardsley1}
p_{f,z} = p_{0,z} + \frac{1}{2 c}\, \vec{A}(\eta_0) \cdot (\vec{A}(\eta_0) - 2\,\vec{p}_{0,\perp}) 
\end{eqnarray}
where $p_{0,z}$ denotes the $z$ component of the electron's initial momentum and $c$ the speed of light \cite{Bardsley_1989} (for details see supplementary information).
Since $|\vec A| \propto \lambda \cdot \sqrt{I}$ the non-dipole contributions can be observed either for high laser intensities $I$ \cite{RMP_2012} or long wavelengths $\lambda$ \cite{Ludwig_2014, Reiss_2008}.
In our experiment, the electron energies are in a regime where the final momentum $p_{f,z} $ in beam direction is different from the initial $p_{0,z}$. 

Besides the two-step model, strong-field ionization can in general be described in a photon picture as multi-photon above-threshold ionization \cite{Agostini_1979}:
A total number of $N_{tot} = N_{sub}+N_{above}$ photons is absorbed and the conserved photon linear momentum $N_{tot} \cdot \hbar k$ is shared between the electron and ion \cite{Eberly_1991}. Whereas the momentum $I_P/c$ of the $N_{sub}$ photons needed to lift the electron into the continuum is rather transferred to the ion \cite{Smeenk_2011}, the quasi-free electron gets accelerated by the remaining $N_{above}$ photons to its final kinetic energy $E_{kin}$ with respect to the atomic core.
Small offsets of the order of ${I_P}/{(3c)}$ in the expectation value of the electron momentum in beam direction from the absorption of the $N_{sub}$ photons have been predicted, but so far not been experimentally confirmed \cite{Klaiber_2013c, Chelkowski_2014}. 

\begin{figure}
\includegraphics{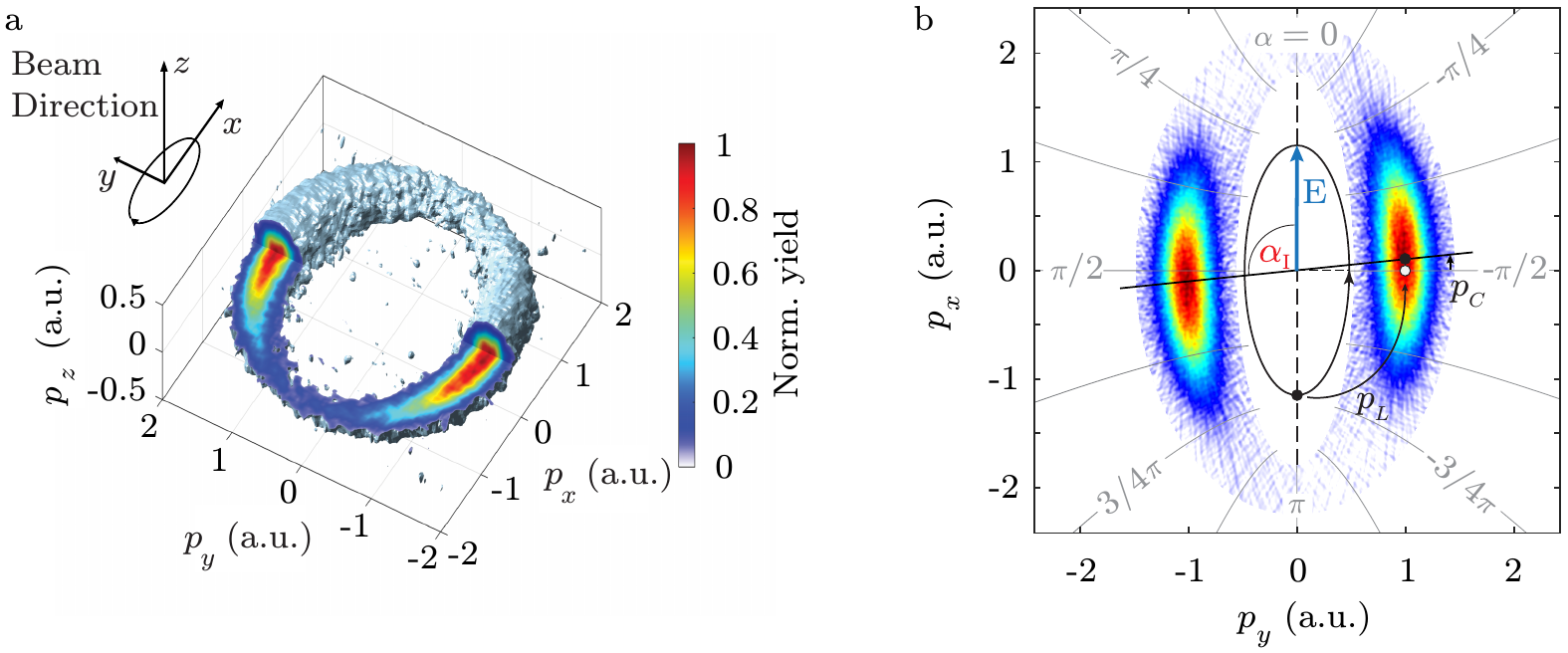}
\caption{\linespread{1.2}\selectfont{}
    \textbf{Momentum distributions from strong-field ionization with angular streaking.}
    ({\bf a}) Isosurface of a reconstructed 3D photoelectron momentum distribution (PMD) recorded at an ellipticity of $\epsilon = 0.8$ together with a sketch of the polarization ellipse and the beam direction. 
    ({\bf b}) Polarization plane PMD for an ellipticity of $\epsilon$ = 0.5 visualizing the definition of elliptical coordinates and the streaking angle $\alpha$. 
    Superimposed is the illustration of the effect of the parent-ion interaction onto an outgoing electron ionized at the peak of the electric field, as indicated by the polarization ellipse in black. The final momentum of the photoelectron gains in addition to the momentum $p_{L}$ acquired by the propagation in the laser field the momentum $p_{C}$ from the interaction with the parent ion mostly in the direction of the instantaneous electric field. The angle $\alpha_I$ denotes the angle of maximal ionization yield.}
\label{Fig2}
\end{figure}

\begin{figure}
    \centering
    \includegraphics[width = 0.495\columnwidth]{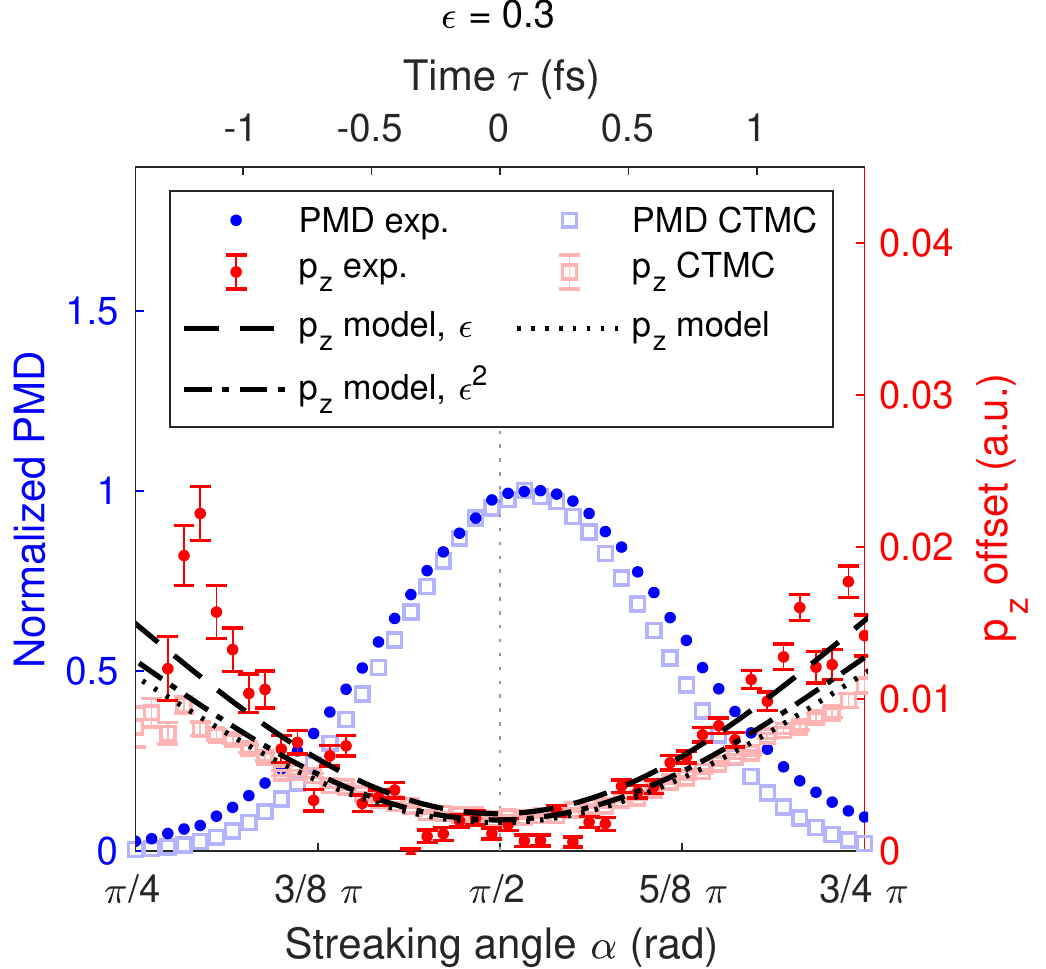}
    \includegraphics[width = 0.495\columnwidth]{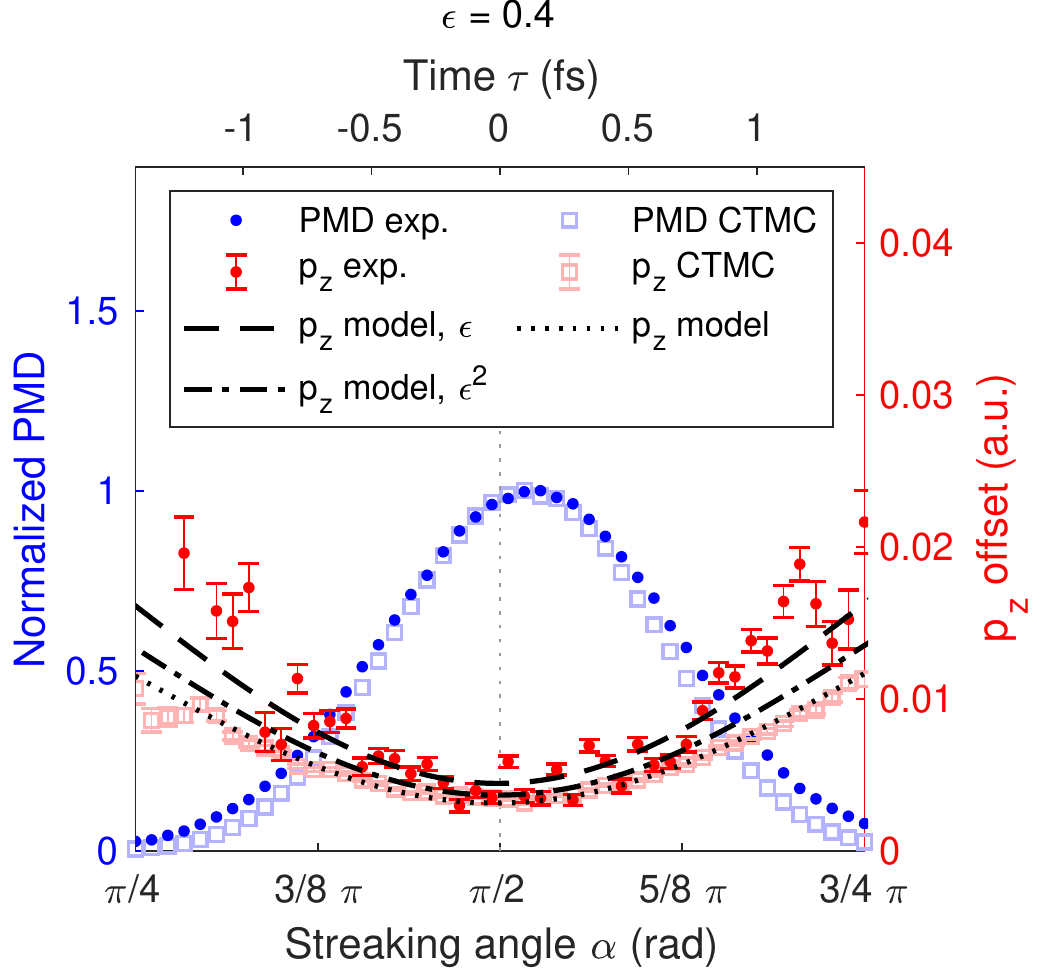}

    \vspace*{1em}
    \includegraphics[width = 0.495\columnwidth]{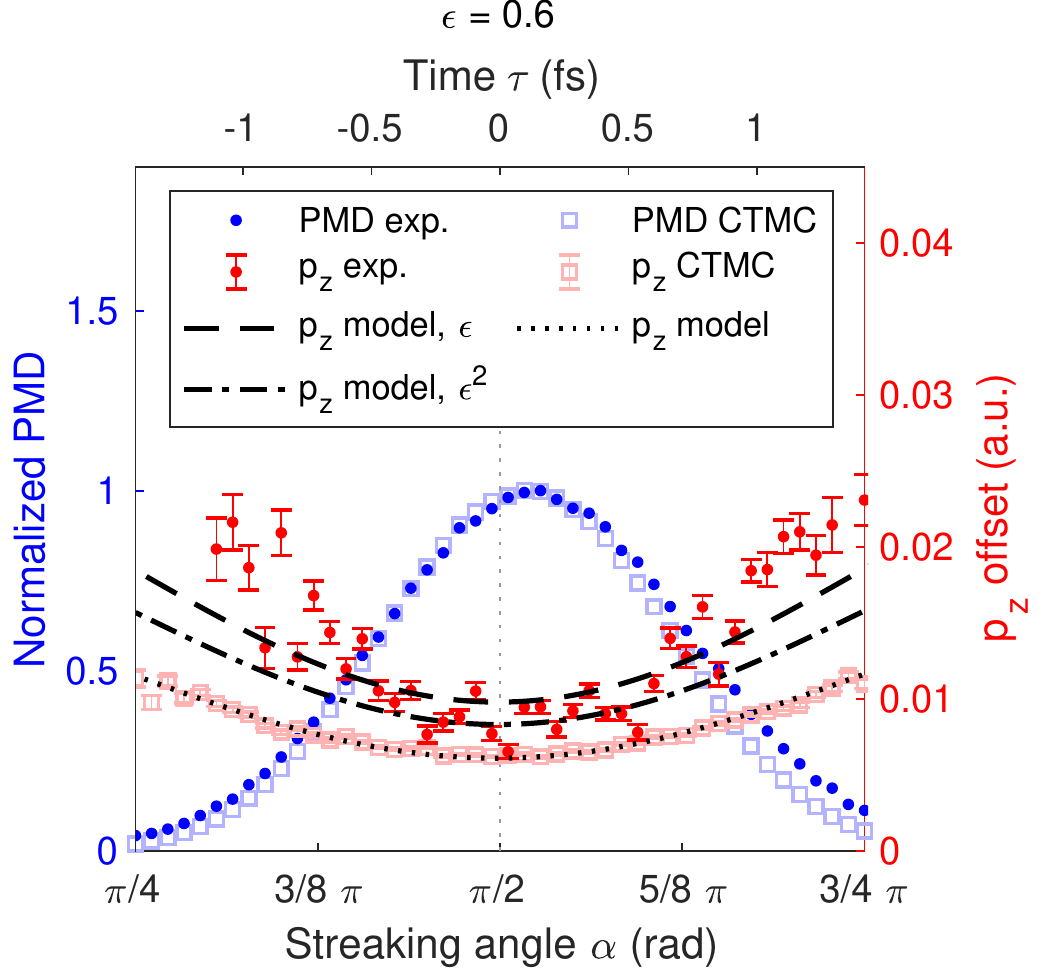}
    \includegraphics[width = 0.495\columnwidth]{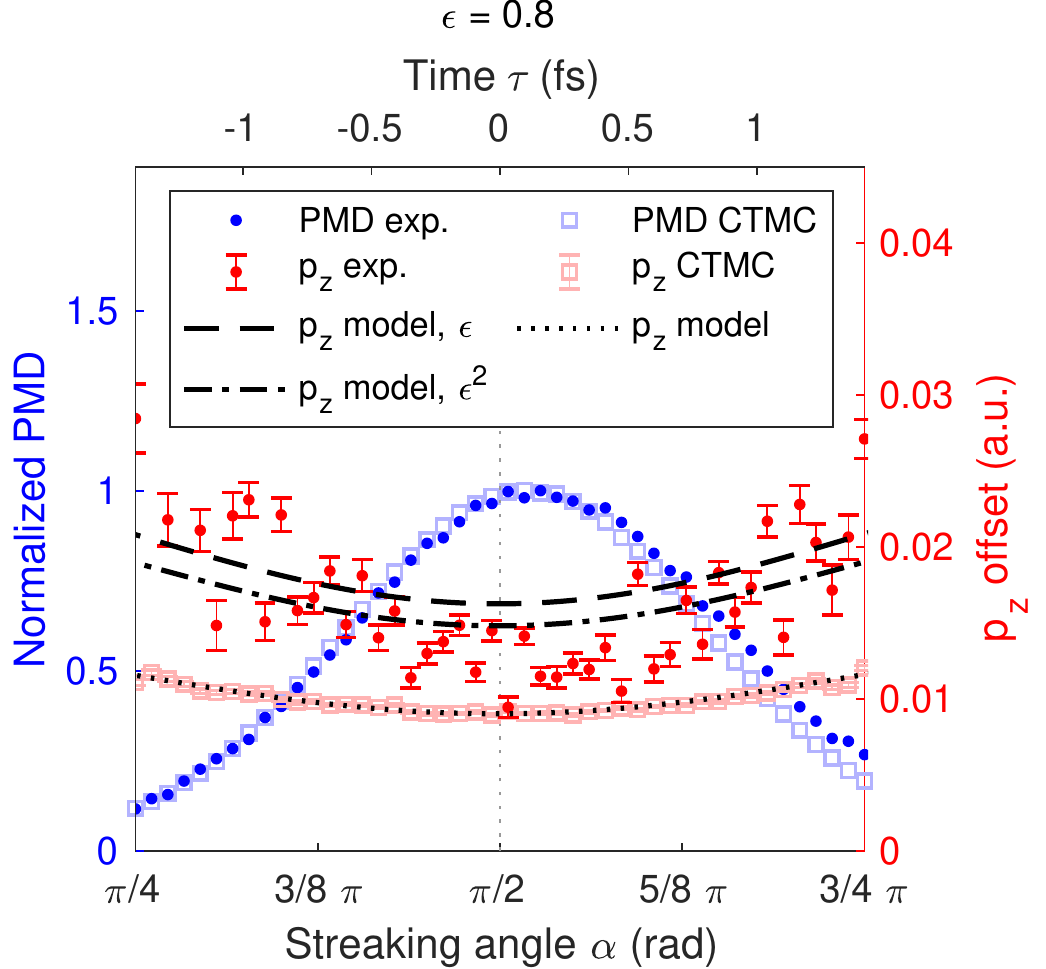}
    \caption{\linespread{1.2}\selectfont{}
    \textbf{Sub-cycle dependence of linear momentum transfer.}
     Angular distributions for various ellipticities of the $p_z$-shift (red) together with the normalized electron yield (blue) from the experimental data (PMD exp., $p_z$ exp.). In addition, we show the corresponding results from our classical trajectory Monte-Carlo (CTMC) simulations (PMD CTMC, $p_z$ CTMC) and the theoretical prediction for a peak intensity of $I_0 = 4\cdot 10^{13}\, \textup{W\,cm}^{-2}$ based on a classical model ("$p_z$ model") and an extended version with an additional initial momentum component along $p_z$ ("$p_z$ model, $\epsilon$" and "$p_z$ model, $\epsilon^2$")(for details see main text and supplementary information). 
    }

    \label{Fig3}
\end{figure}

\begin{figure}
\centering
\includegraphics{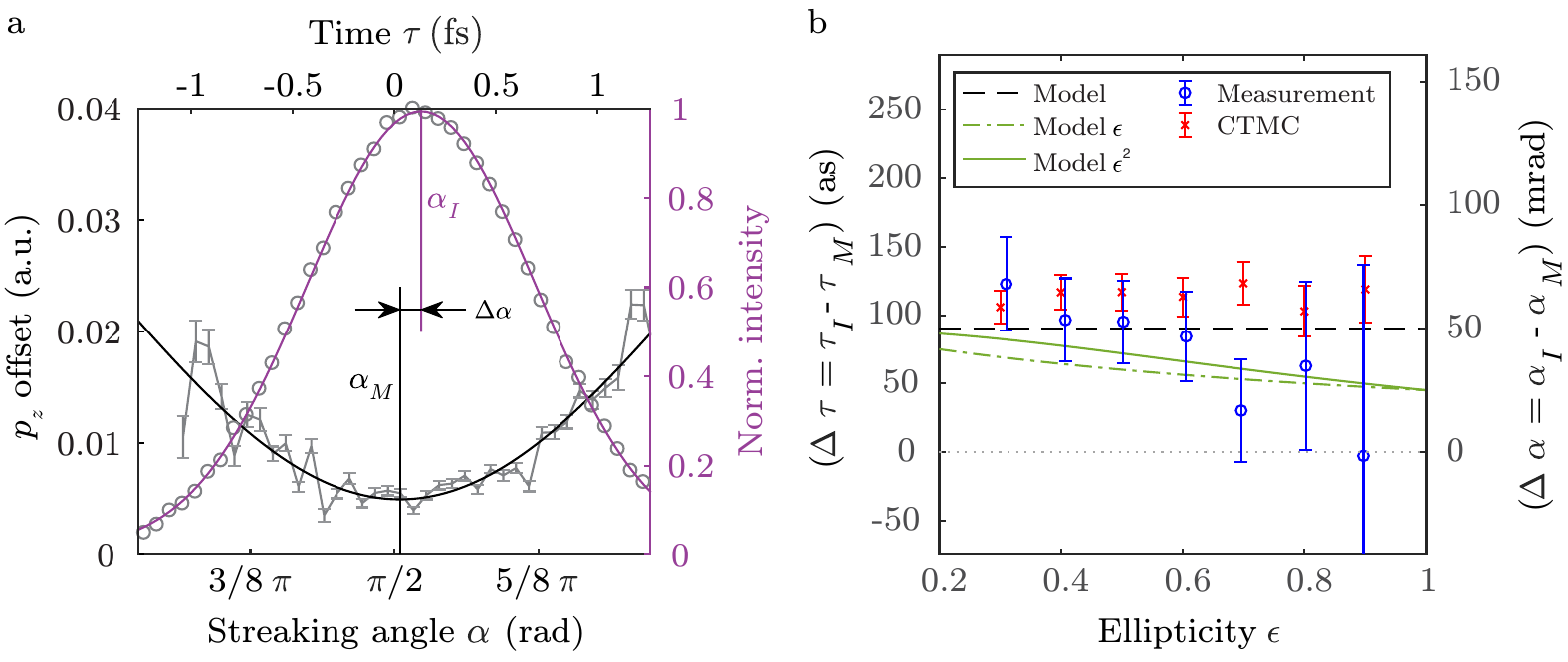}
\caption{\linespread{1.2}\selectfont{}
    \textbf{Attosecond timing of linear momentum transfer.}
    ({\bf a}) Visualization of the angle difference between the angles where $p_z$ minimizes ($\alpha_M$) and the ionization yield maximizes ($\alpha_I$) for an ellipticity of $\epsilon = 0.5$. The corresponding time axis up to a constant shift is shown on top.
    ({\bf b}) Angular difference $\Delta \alpha$ as a function of the ellipticity, together with the corresponding time difference axis. The error bars are based on the 1$\sigma$ uncertainty of the fit. The measurement is compared to the CTMC calculations, and three analytic models for a free electron corrected by the electron-ion interaction ("Model") and extended additionally by an initial momentum along $p_z$ ("Model $\epsilon$" and "Model $\epsilon^2$") (for details see main text and supplementary information). 
}
\label{Fig4}
\end{figure}

In our experiments, we use elliptically polarized mid-infrared (mid-IR) laser pulses centered around $3.4\,\mu$m to strong-field ionize the target xenon in a regime beyond the electric dipole approximation. The laser pulses were compressed based on the ionization signal to $\sim 50$\,fs, corresponding at an optical cycle duration of 11.3\,fs to a pulse duration of 4.4 cycles. The carrier envelope offset phase was not stabilized. At a repetition rate of 50\,kHz we reach a peak intensity of $\sim 4\cdot10^{13}$\, W/cm$^2$. Accordingly, the Keldysh parameter is in the range $\gamma \approx 0.4-0.5$, an intermediate regime between pure multi-photon ionization ($\gamma \gg 1$) and pure tunnel ionization ($\gamma \ll 1$) \cite{Yudin_2001}.

Full 3D photoelectron momentum distributions were recorded with a velocity map imaging spectrometer in combination with a tomographic reconstruction algorithm \cite{Wollenhaupt_2009, Weger_2013} (Fig.~\ref{Fig2}~a). 
The studied ellipticities $\epsilon \geq 0.3$ do not allow for recollisions of the electron with the residual ion \cite{Maurer_2018}.

From the 3D momentum distributions, we extracted the ionization yield as a function of an angle $\alpha$, defined in elliptical coordinates in the polarization plane (Fig.~\ref{Fig2}~b). This ensures a linear mapping of time to angle also for comparably small ellipticites. From the resulting photoelectron angular distribution, we identified an angle $\alpha_I$ where the ionization yield maximizes.
Furthermore, from radially integrated 3D momentum distributions we extracted the shift in $p_z$-direction as a function of $\alpha$ and identified the angle $\alpha_M$ where the momentum transfer from the laser field onto the photoelectrons minimizes.
The details for the data analysis are described in the supplementary information. We observe that the electrons ionized around the maximal electric field 
experience the smallest shift in $p_z$.

In the multi-photon picture of the strong-field ionization process the $p_z$ shift is a direct measure for the number of photons that transferred their linear momentum to the electron. Our measurement for elliptical polarization shows that the number of involved photons varies within an optical cycle. Specifically, for the electrons born around the electric field maximum the number of absorbed photon momenta is smaller than the prediction from the radiation pressure model: Our experiment with ellipticity $\epsilon = 0.5$ shows a minimal $p_z$ shift of $\approx 5\cdot 10^{-3}$\,a.u. corresponding to $\sim$50 photon momenta. The radiation pressure model would predict an average shift of $U_P / c \approx 1\cdot 10^{-2}$\,a.u. corresponding to $\sim$100 photon momenta. $U_P$ is the ponderomotive energy of the photoelectron in the laser field.

This shows that the radiation pressure picture, used to explain the linear momentum transfer in the case of circular polarization \cite{Smeenk_2011}, is not generally applicable for arbitrary polarization states. For elliptical and linear polarization, it is not even generally applicable for the cycle-averaged photelectron momentum $\left\langle p_z \right\rangle$ due to weighting of $p_z$ with the nonlinear ionization rate during the cycle. The radiation pressure picture applies only if the magnetic force onto the electron is constant over one full laser cycle, i.e. for purely circular polarization.

The variation of the $p_z$-shift as a function of the streaking angle $\alpha$, i.e. the ionization phase $\eta_0$, can be largely explained by the previously introduced model for a free electron (equation (\ref{eqn:Bardsley1})). The model predicts a minimum of the $p_z$-shift around ionization phase $\eta_0 = 0$ (see supplementary information) and the correct order of magnitude for the modulation of $\left\langle p_z \right\rangle$ as a function of $\alpha$, including a decrease of the $\left\langle p_z \right\rangle(\alpha)$ variation amplitude with increasing ellipticity (Fig.~\ref{Fig3}).

In the experiment the extracted angles $\alpha_M$ and $\alpha_I$ for minimal $p_z$-shift and maximal PMD signal, respectively, show a positive offset $\Delta \alpha = \alpha_I - \alpha_M$ for all ellipticities (Fig.~\ref{Fig4}, details see supplementary information).
We apply the attoclock principle to translate the angular offset into time. The usage of an elliptical coordinate system in the polarization plane results in a linear mapping between the streaking angle $\alpha$ and the time $\tau$ or ionization phase $\eta_0$ (see supplementary information).
When we translate $\Delta\alpha$ to a time offset $\Delta\tau$, we find that the electrons with a minimal $p_z$-shift are released into the continuum before the maximum of the electric field on the order of 100 as.

Under the assumption of zero initial momentum, the simple model for a free electron predicts a vanishing offset $\Delta \alpha$. 
Both electron trajectories, the most likely one and the one experiencing the smallest linear momentum transfer are ionized at identical phase $\eta_0 = 0$.

The more complete description of the continuum propagation in our classical trajectory Monte-Carlo (CTMC) simulations (see supplementary information) covers the interaction of the photoelectron with the residual ion. As a result the CTMC simulations predict in addition to the ionization phase dependent $\left\langle p_z \right\rangle$ (Fig.~\ref{Fig3}) a negative ionization phase for the minimal $p_z$-shift. This results in a positive $\Delta\alpha$ of the same order of magnitude as found in the experiment (Fig.~\ref{Fig4}(b)).

We include the interaction with the residual ion in the analytic model perturbatively by setting the initial photoelectron momentum to $\vec p_0=\vec p_C$. This is based on the fact that the final momentum can be decomposed in the polarization plane into $\vec{p}_f = \vec{p}_L + \vec{p}_C $, where $\vec{p}_L$ is the momentum transferred onto the electron by the laser field and $\vec{p}_C$ is the momentum acquired by the Coulomb interaction with the residual ion \cite{Goreslavski_2004} (see supplementary information).
The analytic solution for the ionization phase $\eta_0$ with minimal $p_z$,
\begin{eqnarray}
\eta_0 = -0.5\, \arctan\left(\frac{\pi}{\sqrt{2}} \,\omega\, I_P^{-3/2}\right)\,,
\end{eqnarray}
is independent of the laser field intensity and ellipticity $\epsilon$.
The Coulomb corrected analytic predictions are in perfect agreement with the numerical predictions from the CTMC calculations (compare Fig.~\ref{Fig3} and \ref{Fig4}). This shows that the Coulomb interaction is mostly responsible for the angle offset $\Delta\alpha$.

However, the measured absolute values for $\left\langle p_z \right\rangle$ lie slightly above the values expected from both, our CTMC calculations and the analytical model. 
The deviation can be partly caused by the combined influence of the focal volume averaging and an uncertainty in the intensity calibration and the determination of the absolute zero momentum of the detector. The error bars shown in Fig.~\ref{Fig3} are of statistical nature based on the fit and do not include any of the possible systematic errors.
Neither of the above can fully explain the discrepancy in the curvature of the $p_z$-shift as a function of the streaking angle.
Moreover the experimentally measured delay times $\Delta\tau$ appear to decrease with increasing ellipticity, whereas the CTMC-simulations and the analytic model predict a constant value (Fig.~\ref{Fig4}(b)).

We suggest to introduce an additional initial momentum along the beam propagation direction of $\frac{1}{2c}f(\epsilon)|\vec A(\eta)|^2$, where $f(\epsilon)$ is a non constant function of the ellipticity $\epsilon$. This correction is based on the momentum corresponding to the additional energy of the electron in the laser field described by the vector potential $A(\eta)$ when it appears at the phase $\eta$ in the continuum (see supplementary information). 
Extended to the total initial momentum $\vec p_0 = \vec p_C + \frac{1}{2c}f(\epsilon)|\vec A(\eta)|^2 \hat k$ the analytic model predicts $\left\langle p_z \right\rangle(\alpha)$ curves following more closely the measured ones (Fig.~\ref{Fig3}). Also, the ionization phase $\eta_0$ corresponding to minimal momentum transfer $p_z$,
\begin{eqnarray}
\eta_0 = -0.5 \arctan\left(\frac{\pi}{\sqrt{2} (1+f(\epsilon))} \,\omega\, I_P^{-3/2}\right) \,,
\end{eqnarray}
becomes ellipticity dependent. Our model shows good agreement for the two functions $f(\epsilon) = \epsilon$ and $f(\epsilon) = \epsilon^2$ with the experimental results (Fig.~\ref{Fig4}(b)).

In conclusion, we experimentally demonstrated that the strong-field momentum transfer in laser propagation direction from the field onto the photoelectrons beyond the limit of the dipole approximation is a time-dependent process within an optical cycle.
Thus, a time-averaged radiation pressure picture is not applicable in the general case of elliptical polarization. 
The time dependent momentum transfer can mostly be explained with a classical model of a free electron, extended by the parent ion interaction of the escaping photoelectron and an additional time-dependent initial momentum shift related to the energy of an electron in an electromagnetic field at the phase of ionization $\eta_0$.
The shift of the PMD along the laser propagation direction provides direct access to the number of photons that were absorbed by the electron during the ionization process. Hereby, our results open up new possibilities for measurements on the timing of photon absorption as well as fluctuations of the laser field \cite{Riek_2017}.

Furthermore, we observed a time delay between the times of maximal ionization yield and minimal linear momentum transfer. As the ionization rate is connected to the cycles of the field, our observations imply a time delay between the cycles of the field and the linear momentum transfer. 
We showed that the electrons with the smallest momentum transfer are ionized shortly before the peak of the electric field with a time delay on the order of several tens of attoseconds (Fig. \ref{Fig4}). 
This time delay might contain further information about the electron dynamics in the classically forbidden region during the ionization process. 

Our findings have important consequences for all areas of physics that are influenced by the field-momentum transfer. They also trigger the question about time delays in the field-momentum transfer in the case of single photon ionization, i.e. if the parent ion potential induced time delays are a general property of photoionization or apply only for the case of strong field ionization. Thus, our results can motivate further studies with single photon ionization via time-dependent $p_z$-transfer in streaking- and RABBITT-experiments.

{\bf Data availability.} The data that support the plots within this paper and other findings of this study are available from the corresponding author upon reasonable request. 

\bibliography{AttoPz}

\begin{thebibliography}{10}
\expandafter\ifx\csname url\endcsname\relax
  \def\url#1{\texttt{#1}}\fi
\expandafter\ifx\csname urlprefix\endcsname\relax\def\urlprefix{URL }\fi
\providecommand{\bibinfo}[2]{#2}
\providecommand{\eprint}[2][]{\url{#2}}

\bibitem{Willenberg_17}
\bibinfo{author}{Willenberg, B.} \emph{et~al.}
\newblock \bibinfo{title}{Sub-cycle resolution of field-momentum transfer in
  non-dipole strong-field ionization}.
\newblock In \emph{\bibinfo{booktitle}{CLEO EU 17}}, \bibinfo{pages}{CG\_7\_4}
  (\bibinfo{publisher}{Optical Society of America}, \bibinfo{year}{2017}).

\bibitem{Willenberg_18}
\bibinfo{author}{Willenberg, B.}, \bibinfo{author}{Maurer, J.},
  \bibinfo{author}{Mayer, B.~W.} \& \bibinfo{author}{Keller, U.}
\newblock \bibinfo{title}{Linear momentum transfer in multiphoton strong-field
  ionization with subcycle time resolution}.
\newblock In \emph{\bibinfo{booktitle}{AttoFEL 18}} (\bibinfo{address}{London},
  \bibinfo{year}{2018}).

\bibitem{Maurer_18}
\bibinfo{author}{Willenberg, B.}, \bibinfo{author}{Maurer, J.},
  \bibinfo{author}{Mayer, B.~W.} \& \bibinfo{author}{Keller, U.}
\newblock \bibinfo{title}{Linear momentum transfer in multiphoton strong-field
  ionization with subcycle time resolution}.
\newblock In \emph{\bibinfo{booktitle}{LPHYS 18}}
  (\bibinfo{address}{Nottingham}, \bibinfo{year}{2018}).

\bibitem{Dorner_2000}
\bibinfo{author}{D{\"o}rner, R.} \emph{et~al.}
\newblock \bibinfo{title}{Cold target recoil ion momentum spectroscopy: a
  ‘momentum microscope’ to view atomic collision dynamics}.
\newblock \emph{\bibinfo{journal}{Physics Reports}}
  \textbf{\bibinfo{volume}{330}}, \bibinfo{pages}{95 -- 192}
  (\bibinfo{year}{2000}).

\bibitem{Esirkepov_2004}
\bibinfo{author}{Esirkepov, T.}, \bibinfo{author}{Borghesi, M.},
  \bibinfo{author}{Bulanov, S.~V.}, \bibinfo{author}{Mourou, G.} \&
  \bibinfo{author}{Tajima, T.}
\newblock \bibinfo{title}{Highly efficient relativistic-ion generation in the
  laser-piston regime}.
\newblock \emph{\bibinfo{journal}{Phys. Rev. Lett.}}
  \textbf{\bibinfo{volume}{92}}, \bibinfo{pages}{175003}
  (\bibinfo{year}{2004}).

\bibitem{Pegoraro_2007}
\bibinfo{author}{Pegoraro, F.} \& \bibinfo{author}{Bulanov, S.}
\newblock \bibinfo{title}{Photon bubbles and ion acceleration in a plasma
  dominated by the radiation pressure of an electromagnetic pulse}.
\newblock \emph{\bibinfo{journal}{Phys. Rev. Lett.}}
  \textbf{\bibinfo{volume}{99}}, \bibinfo{pages}{065002}
  (\bibinfo{year}{2007}).

\bibitem{Wineland_1978}
\bibinfo{author}{Wineland, D.~J.}, \bibinfo{author}{Drullinger, R.~E.} \&
  \bibinfo{author}{Walls, F.~L.}
\newblock \bibinfo{title}{Radiation-pressure cooling of bound resonant
  absorbers}.
\newblock \emph{\bibinfo{journal}{Phys. Rev. Lett.}}
  \textbf{\bibinfo{volume}{40}}, \bibinfo{pages}{1639--1642}
  (\bibinfo{year}{1978}).

\bibitem{Aspect_1988}
\bibinfo{author}{Aspect, A.}, \bibinfo{author}{Arimondo, E.},
  \bibinfo{author}{Kaiser, R.}, \bibinfo{author}{Vansteenkiste, N.} \&
  \bibinfo{author}{Cohen-Tannoudji, C.}
\newblock \bibinfo{title}{Laser cooling below the one-photon recoil energy by
  velocity-selective coherent population trapping}.
\newblock \emph{\bibinfo{journal}{Phys. Rev. Lett.}}
  \textbf{\bibinfo{volume}{61}}, \bibinfo{pages}{826--829}
  (\bibinfo{year}{1988}).

\bibitem{Gigan_2006}
\bibinfo{author}{Gigan, S.} \emph{et~al.}
\newblock \bibinfo{title}{Self-cooling of a micromirror by radiation pressure}.
\newblock \emph{\bibinfo{journal}{Nature}} \textbf{\bibinfo{volume}{444}},
  \bibinfo{pages}{67} (\bibinfo{year}{2006}).

\bibitem{Landau_4}
\bibinfo{author}{Berestetskii, V.~B.}, , \bibinfo{author}{Lifshitz, E.~M.} \&
  \bibinfo{author}{Pitevskii, L.~P.}
\newblock \emph{\bibinfo{title}{Quantum electrodynamics}}
  (\bibinfo{publisher}{Pergamon, Oxford}, \bibinfo{year}{1982}).

\bibitem{Sommerfeld_1930}
\bibinfo{author}{Sommerfeld, A.} \& \bibinfo{author}{Schur, G.}
\newblock \bibinfo{title}{{\"U}ber den photoeffekt in der k-schale der atome,
  insbesondere {\"u}ber die voreilung der photoelektronen}.
\newblock \emph{\bibinfo{journal}{Annalen der Physik}}
  \textbf{\bibinfo{volume}{396}}, \bibinfo{pages}{409--432}
  (\bibinfo{year}{1930}).

\bibitem{schiff1955quantum}
\bibinfo{author}{Schiff, L.}
\newblock \emph{\bibinfo{title}{Quantum Mechanics}}.
\newblock International series in pure and applied physics
  (\bibinfo{publisher}{McGraw-Hill}, \bibinfo{year}{1955}).

\bibitem{Michaud_1970}
\bibinfo{author}{Michaud, G.}
\newblock \bibinfo{title}{Diffusion processes in peculiar a stars}.
\newblock \emph{\bibinfo{journal}{The Astrophysical Journal}}
  \textbf{\bibinfo{volume}{160}}, \bibinfo{pages}{641} (\bibinfo{year}{1970}).

\bibitem{Seaton_1995}
\bibinfo{author}{Seaton, M.}
\newblock \bibinfo{title}{Momentum transfer in photo-ionization processes}.
\newblock \emph{\bibinfo{journal}{J. Phys. B: Atomic, Molecular and Optical
  Physics}} \textbf{\bibinfo{volume}{28}}, \bibinfo{pages}{3185}
  (\bibinfo{year}{1995}).

\bibitem{Massacrier_1996}
\bibinfo{author}{Massacrier, G.}
\newblock \bibinfo{title}{Momentum transfer in photoionization of one-electron
  ions}.
\newblock \emph{\bibinfo{journal}{Astronomy and Astrophysics}}
  \textbf{\bibinfo{volume}{309}}, \bibinfo{pages}{979--990}
  (\bibinfo{year}{1996}).

\bibitem{Chelkowski_2014}
\bibinfo{author}{Chelkowski, S.}, \bibinfo{author}{Bandrauk, A.~D.} \&
  \bibinfo{author}{Corkum, P.~B.}
\newblock \bibinfo{title}{Photon momentum sharing between an electron and an
  ion in photoionization: From one-photon (photoelectric effect) to multiphoton
  absorption}.
\newblock \emph{\bibinfo{journal}{Phys. Rev. Lett.}}
  \textbf{\bibinfo{volume}{113}}, \bibinfo{pages}{263005}
  (\bibinfo{year}{2014}).

\bibitem{Moore_1995}
\bibinfo{author}{Moore, C.~I.}, \bibinfo{author}{Knauer, J.~P.} \&
  \bibinfo{author}{Meyerhofer, D.~D.}
\newblock \bibinfo{title}{Observation of the transition from thomson to compton
  scattering in multiphoton interactions with low-energy electrons}.
\newblock \emph{\bibinfo{journal}{Phys. Rev. Lett.}}
  \textbf{\bibinfo{volume}{74}}, \bibinfo{pages}{2439--2442}
  (\bibinfo{year}{1995}).

\bibitem{Smeenk_2011}
\bibinfo{author}{Smeenk, C. T.~L.} \emph{et~al.}
\newblock \bibinfo{title}{Partitioning of the linear photon momentum in
  multiphoton ionization}.
\newblock \emph{\bibinfo{journal}{Phys. Rev. Lett.}}
  \textbf{\bibinfo{volume}{106}}, \bibinfo{pages}{193002}
  (\bibinfo{year}{2011}).

\bibitem{Ludwig_2014}
\bibinfo{author}{Ludwig, A.} \emph{et~al.}
\newblock \bibinfo{title}{Breakdown of the dipole approximation in strong-field
  ionization}.
\newblock \emph{\bibinfo{journal}{Phys. Rev. Lett.}}
  \textbf{\bibinfo{volume}{113}}, \bibinfo{pages}{243001}
  (\bibinfo{year}{2014}).

\bibitem{Chelkowski_2015}
\bibinfo{author}{Chelkowski, S.}, \bibinfo{author}{Bandrauk, A.~D.} \&
  \bibinfo{author}{Corkum, P.~B.}
\newblock \bibinfo{title}{Photon-momentum transfer in multiphoton ionization
  and in time-resolved holography with photoelectrons}.
\newblock \emph{\bibinfo{journal}{Phys. Rev. A}} \textbf{\bibinfo{volume}{92}},
  \bibinfo{pages}{051401} (\bibinfo{year}{2015}).

\bibitem{Liu_2013}
\bibinfo{author}{Liu, J.}, \bibinfo{author}{Xia, Q.~Z.}, \bibinfo{author}{Tao,
  J.~F.} \& \bibinfo{author}{Fu, L.~B.}
\newblock \bibinfo{title}{Coulomb effects in photon-momentum partitioning
  during atomic ionization by intense linearly polarized light}.
\newblock \emph{\bibinfo{journal}{Phys. Rev. A}} \textbf{\bibinfo{volume}{87}},
  \bibinfo{pages}{041403} (\bibinfo{year}{2013}).

\bibitem{Maurer_2018}
\bibinfo{author}{Maurer, J.} \emph{et~al.}
\newblock \bibinfo{title}{Probing the ionization wave packet and recollision
  dynamics with an elliptically polarized strong laser field in the nondipole
  regime}.
\newblock \emph{\bibinfo{journal}{Phys. Rev. A}} \textbf{\bibinfo{volume}{97}},
  \bibinfo{pages}{013404} (\bibinfo{year}{2018}).

\bibitem{Eckle_2008a}
\bibinfo{author}{Eckle, P.} \emph{et~al.}
\newblock \bibinfo{title}{Attosecond angular streaking}.
\newblock \emph{\bibinfo{journal}{Nat. Phys.}} \textbf{\bibinfo{volume}{4}},
  \bibinfo{pages}{565} (\bibinfo{year}{2008}).

\bibitem{Eckle_2008b}
\bibinfo{author}{Eckle, P.} \emph{et~al.}
\newblock \bibinfo{title}{Attosecond ionization and tunneling delay time
  measurements in helium}.
\newblock \emph{\bibinfo{journal}{Science}} \textbf{\bibinfo{volume}{322}},
  \bibinfo{pages}{1525--1529} (\bibinfo{year}{2008}).

\bibitem{Telle1999}
\bibinfo{author}{Telle, H.} \emph{et~al.}
\newblock \bibinfo{title}{Carrier-envelope offset phase control: A novel
  concept for absolute optical frequency measurement and ultrashort pulse
  generation}.
\newblock \emph{\bibinfo{journal}{Applied Physics B}}
  \textbf{\bibinfo{volume}{69}}, \bibinfo{pages}{327--332}
  (\bibinfo{year}{1999}).

\bibitem{Sarachik_1970}
\bibinfo{author}{Sarachik, E.} \& \bibinfo{author}{Schappert, G.}
\newblock \bibinfo{title}{Classical theory of the scattering of intense laser
  radiation by free electrons}.
\newblock \emph{\bibinfo{journal}{Phys. Rev. D}} \textbf{\bibinfo{volume}{1}},
  \bibinfo{pages}{2738} (\bibinfo{year}{1970}).

\bibitem{TwoStep}
\bibinfo{howpublished}{see e.g. H. B. van Linden van den Heuvell and H. G.
  Muller, in Multiphoton Processes, ed. S. J. Smith and P. L. Knight, Cambridge
  University Press (1988); T.F. Gallagher, Phys. Rev. Lett. {\bf 61}, 2304
  (1988); K. J. Schafer, B. Yang, L. F. DiMauro, and K. C. Kulander, Phys. Rev.
  Lett. {\bf 70}, 1599 (1993); P. B. Corkum, Phys. Rev. Lett. {\bf 71}, 1994
  (1993)}.

\bibitem{Bardsley_1989}
\bibinfo{author}{Bardsley, J.~N.}, \bibinfo{author}{Penetrante, B.~M.} \&
  \bibinfo{author}{Mittleman, M.~H.}
\newblock \bibinfo{title}{Relativistic dynamics of electrons in intense laser
  fields}.
\newblock \emph{\bibinfo{journal}{Phys. Rev. A}} \textbf{\bibinfo{volume}{40}},
  \bibinfo{pages}{3823--3835} (\bibinfo{year}{1989}).

\bibitem{RMP_2012}
\bibinfo{author}{Di~Piazza, A.}, \bibinfo{author}{M\"uller, C.},
  \bibinfo{author}{Hatsagortsyan, K.~Z.} \& \bibinfo{author}{Keitel, C.~H.}
\newblock \bibinfo{title}{Extremely high-intensity laser interactions with
  fundamental quantum systems}.
\newblock \emph{\bibinfo{journal}{Rev. Mod. Phys.}}
  \textbf{\bibinfo{volume}{84}}, \bibinfo{pages}{1177--1228}
  (\bibinfo{year}{2012}).

\bibitem{Reiss_2008}
\bibinfo{author}{Reiss, H.~R.}
\newblock \bibinfo{title}{Limits on tunneling theories of strong-field
  ionization}.
\newblock \emph{\bibinfo{journal}{Phys. Rev. Lett.}}
  \textbf{\bibinfo{volume}{101}}, \bibinfo{pages}{043002}
  (\bibinfo{year}{2008}).

\bibitem{Agostini_1979}
\bibinfo{author}{Agostini, P.}, \bibinfo{author}{Fabre, F.},
  \bibinfo{author}{Mainfray, G.}, \bibinfo{author}{Petite, G.} \&
  \bibinfo{author}{Rahman, N.~K.}
\newblock \bibinfo{title}{Free-free transitions following six-photon ionization
  of xenon atoms}.
\newblock \emph{\bibinfo{journal}{Phys. Rev. Lett.}}
  \textbf{\bibinfo{volume}{42}}, \bibinfo{pages}{1127--1130}
  (\bibinfo{year}{1979}).

\bibitem{Eberly_1991}
\bibinfo{author}{Eberly, J.~H.}, \bibinfo{author}{Javanainen, J.} \&
  \bibinfo{author}{Rz\k{a}{\.z}ewski, K.}
\newblock \bibinfo{title}{Above-threshold ionization}.
\newblock \emph{\bibinfo{journal}{Physics reports}}
  \textbf{\bibinfo{volume}{204}}, \bibinfo{pages}{331--383}
  (\bibinfo{year}{1991}).

\bibitem{Klaiber_2013c}
\bibinfo{author}{Klaiber, M.}, \bibinfo{author}{Yakaboylu, E.},
  \bibinfo{author}{Bauke, H.}, \bibinfo{author}{Hatsagortsyan, K.~Z.} \&
  \bibinfo{author}{Keitel, C.~H.}
\newblock \bibinfo{title}{Under-the-barrier dynamics in laser-induced
  relativistic tunneling}.
\newblock \emph{\bibinfo{journal}{Phys. Rev. Lett.}}
  \textbf{\bibinfo{volume}{110}}, \bibinfo{pages}{153004}
  (\bibinfo{year}{2013}).

\bibitem{ADK}
\bibinfo{author}{Ammosov, M.~V.}, \bibinfo{author}{Delone, N.~B.} \&
  \bibinfo{author}{Krainov, V.~P.}
\newblock \bibinfo{title}{Tunnel ionization of complex atoms and of atomic ions
  in an alternating electromagnetic field}.
\newblock \emph{\bibinfo{journal}{Zh. Eksp. Teor. Fiz.}}
  \textbf{\bibinfo{volume}{91}}, \bibinfo{pages}{2008} (\bibinfo{year}{1986}).

\bibitem{Yudin_2001}
\bibinfo{author}{Yudin, G.~L.} \& \bibinfo{author}{Ivanov, M.~Y.}
\newblock \bibinfo{title}{Nonadiabatic tunnel ionization: Looking inside a
  laser cycle}.
\newblock \emph{\bibinfo{journal}{Phys. Rev. A}} \textbf{\bibinfo{volume}{64}},
  \bibinfo{pages}{013409} (\bibinfo{year}{2001}).

\bibitem{Wollenhaupt_2009}
\bibinfo{author}{Wollenhaupt, M.} \emph{et~al.}
\newblock \bibinfo{title}{Three-dimensional tomographic reconstruction of
  ultrashort free electron wave packets}.
\newblock \emph{\bibinfo{journal}{Appl. Phys. B}}
  \textbf{\bibinfo{volume}{95}}, \bibinfo{pages}{647--651}
  (\bibinfo{year}{2009}).

\bibitem{Weger_2013}
\bibinfo{author}{Weger, M.}, \bibinfo{author}{Maurer, J.},
  \bibinfo{author}{Ludwig, A.}, \bibinfo{author}{Gallmann, L.} \&
  \bibinfo{author}{Keller, U.}
\newblock \bibinfo{title}{Transferring the attoclock technique to velocity map
  imaging}.
\newblock \emph{\bibinfo{journal}{Opt. Expr.}} \textbf{\bibinfo{volume}{21}},
  \bibinfo{pages}{21981--21990} (\bibinfo{year}{2013}).

\bibitem{Goreslavski_2004}
\bibinfo{author}{Goreslavski, S.~P.}, \bibinfo{author}{Paulus, G.~G.},
  \bibinfo{author}{Popruzhenko, S.~V.} \& \bibinfo{author}{Shvetsov-Shilovski,
  N.~I.}
\newblock \bibinfo{title}{Coulomb asymmetry in above-threshold ionization}.
\newblock \emph{\bibinfo{journal}{Phys. Rev. Lett.}}
  \textbf{\bibinfo{volume}{93}}, \bibinfo{pages}{233002}
  (\bibinfo{year}{2004}).

\bibitem{Riek_2017}
\bibinfo{author}{Riek, C.} \emph{et~al.}
\newblock \bibinfo{title}{Subcycle quantum electrodynamics}.
\newblock \emph{\bibinfo{journal}{Nature}} \textbf{\bibinfo{volume}{541}},
  \bibinfo{pages}{376--379} (\bibinfo{year}{2017}).

\end{thebibliography}


\begin{addendum}
 \item This research was supported by the NCCR MUST, funded by the Swiss National Science Foundation and by the ERC advanced grant  ERC ADG AttoClock - 320401 within the seventh framework programme of the European Union. B.W. was supported by an ETH Research Grant ETH-11~15-1.
 \item[Author contributions] B.W., J.M. and B.W.M. performed the experiment. B.W. and J.M. carried out the data analysis and numerical simulations. B.W., J.M. and U.K. interpreted the data and the simulations. All authors contributed to writing the manuscript.
 \item[Additional information] ~\\
    {\bf Competing Interests:} The authors declare that they have no competing financial interests. \\
    {\bf Correspondence and requests for materials} should be addressed to B.W. or U.K.. \\
\end{addendum}

\end{document}